\documentclass{jfm}
\usepackage{graphicx}
\usepackage{epstopdf, epsfig}
\usepackage{natbib}
\usepackage[colorlinks=true,citecolor=blue]{hyperref}
\usepackage{float}
\usepackage{tikz}
\usepackage{soul}
\usepackage{makecell}

\usepackage{lineno}

\shorttitle{Multiscale coherent dynamics in wind turbine wakes}
\shortauthor{N. Biswas, and O.R.H. Buxton}
\title{Effect of tip speed ratio on multiscale coherent dynamics in the near wake of a model wind turbine
}

\author{Neelakash Biswas\aff{1} \corresp{\email{n.biswas20@imperial.ac.uk}},
  Oliver R.H. Buxton \aff{1}
}

\affiliation{\aff{1}Department of Aeronautics, Imperial College London,
London, SW7 2AZ, UK

}

\graphicspath{{figures/}}

\newcommand{\NB}[1]{{\textcolor{black}{#1}}} 

\begin{document}

\maketitle

\begin{abstract}
The near wake of a small-scale wind turbine is investigated using Particle Image Velocimetry (PIV) experiments at different tip speed ratios ($\lambda$). The wind turbine model had a nacelle and a tower mimicking real scale wind turbines. The near wake is found to be dominated by multiple coherent structures including the tip vortices, distinct vortex sheddings from the nacelle and tower, and wake meandering. The merging of the tip vortices is found to be strongly dependent on $\lambda$. A convective length scale ($L_c$) related to the pitch of the helical vortices is defined which is shown to be a better length scale than turbine diameter ($D$) to demarcate the near wake from the far wake. The tower induced strong vertical asymmetry in the flow by destabilising the tip vortices and promoting mixing in the lower (below the nacelle) plane. The nacelle's shedding is found to be important in `seeding' wake meandering which although not potent, exists close to the nacelle, and it becomes important only after a certain distance downstream ($x>3L_c$). A link between the `effective porosity' of the turbine and $\lambda$ is established and the strength and frequency of wake meandering is found to be dependent on $\lambda$. In fact, a decreasing trend of wake meandering frequency with $\lambda$ is observed, similar to vortex shedding from a porous disk at varying porosity. Such similarity upholds the notion of wake meandering being a global instability of the turbine, which can be considered as a `porous' bluff body of diameter $D$.

\end{abstract}


\section{Introduction}
Over recent years, the world has seen a tremendous growth in the wind energy sector with the global wind energy production capacity in 2021 standing nearly 4.5 times the capacity in 2010 \citep{IEA}. However, aligning with ambitious net-zero targets still requires a colossal effort to exploit the full potential of wind energy resources through improved design of wind farm layouts as well as individual turbines. The performance of a single wind turbine can be affected by several factors such as interaction with the atmospheric boundary layer, inflow turbulence, presence of complex topologies, atmospheric stability and so on. In addition, inside a grid of turbines, the wake produced from one turbine can contribute to significant power losses and fatigue damage to subsequent wind turbines downstream \citep{vermeer2003wind, sanderse2011review, stevens2017flow, porte2020wind}. Therefore, a better understanding of the spatial development of a turbine-wake, as well as its dynamic properties, is necessary. {With ever increasing turbine diameter, particularly for offshore turbines, the turbine spacing is no longer a free parameter that can be solely decided based on optimising for total power output, rather land/area related constraints also become a key factor in designing wind farm layouts \citep{lignarolo2016experimental, gaumond2012benchmarking, howland2019wind}. In that regard, studying the near wake of the turbine, where strong coherence is present, becomes particularly important. } 

The near wake (coarsely defined as the region within 3 rotor diameters (D) downstream of the rotor plane) of a wind turbine is multiscale in nature as the flow is forced simultaneously at multiple length scales, for example by the tower, nacelle, blade tip/root vortices etc., thereby introducing coherence into the overall wake at multiple time/length scales \citep{porte2020wind, crespo1999survey, abraham2019effect}. Out of these, the most pronounced structures in the near wake are the tip vortices, the dynamics of which has been studied extensively through numerous experimental \citep{sherry2013interaction, okulov2014regular, lignarolo2014experimental, lignarolo2015tip} and numerical studies \citep{ivanell2010stability, lu2011large, sarmast2014mutual, hodgkin2022numerical}. The stability analysis of \citet{widnall1972stability} showed the presence of three different instability modes of the tip vortices, a short-wave mode, a long-wave mode and the mutual inductance mode, out of which the mutual inductance mode was found to dominate the breakup process of the helicoidal vortex system. The intensity of the mode depends on the pitch of the helical vortex system, hence on the tip speed ratio ($\lambda = \omega R/U_{\infty}$, where $R$ is the turbine radius, $\omega$ is the rotational speed, and $U_{\infty}$ is the free stream velocity) \citep{sherry2013interaction, okulov2014regular}. Depending on the blade configuration, distinct root vortices can also form near the root region of the blade \citep{sherry2013interaction}. However, they are much less persistent in comparison to the tip vortices, whose dynamics and breakdown is particularly important to initiate the recovery of the turbine wake. \citet{medici2005experimental} noted that the tip vortex system in the near field acts as a shield, restricting the exchange of mass and momentum with the outer, background fluid. In other words, breakdown of the tip vortices is a necessary process to re-energise the wake in the far field, reducing the velocity deficit, which is beneficial for the subsequent turbines in the grid. 

Interestingly, \citet{okulov2014regular} found a single dominant frequency in the far wake (which they defined as streamwise distance, $x > 2.5D$) of a wind turbine which was nearly independent of operating conditions. The corresponding Strouhal number based on rotor diameter and freestream velocity was 0.23. A similar Strouhal number in the range (0.15 - 0.4) has been noted in several other works \citep{chamorro2013interaction, foti2018wake, medici2008measurements, heisel2018spectral}. This Strouhal number was associated with wake meandering which is responsible for large transverse displacements of the wake centreline (which can be roughly defined as the location of maximum velocity deficit) in the far wake of the turbine. Although, the dominance of wake meandering in the far wake has been known for many years, the scientific community still holds varied opinions about the genesis of the meandering motion in the far wake. For instance, wake meandering has been seen as a passive advection of the turbine wake due to large scale atmospheric structures \citep{larsen2008wake, espana2011spatial}. In contrast, \citet{okulov2014regular} observed wake meandering when the free stream turbulence level was negligible. The authors proposed that wake meandering could be related to the instability of the shed vortices and connected it to the slow precession motion of the helicoidal vortex system. On a similar note, the importance of the nacelle in the generation of wake meandering was reported through linear stability analysis \citep{iungo2013linear} and numerical simulations \citep{kang2014onset, foti2016wake, foti2018wake}. The large eddy simulations of \citet{foti2018wake} showed that wake meandering is related to the slow precession motion of the helical hub vortex formed behind the nacelle. The authors suggested that the hub vortex grows radially and interacts with the outer wake which can potentially augment wake meandering. Even after a significant amount of research, the exact cause of wake meandering still remains elusive which is perhaps indicated by the ineludible scatter of Strouhal numbers associated with wake meandering observed in different studies \citep{medici2008measurements}. The distance from the rotor plane, where a wake meandering frequency has been observed, has also varied in different studies. For instance, \citet{chamorro2013interaction} found wake meandering only after 3 rotor diameters whereas, \citet{okulov2014regular} reported the presence of a wake meandering frequency as close as $1.5D$ from the rotor (see fig. 6 of \citet{okulov2014regular}) where they found small variation in the frequency for different operating conditions. \citet{medici2008measurements} reported a similar observation at 1 diameters dowstream of the rotor and concluded that wake meandering frequency varied with both tip speed ratio and the thrust coefficient of the rotor. In the present study we attempt to address some of these discrepancies through extensive laboratory experiments on a model wind turbine. 

We perform a series of particle image velocimetry (PIV) experiments to study the near wake of a wind turbine at a range of tip speed ratios, while focusing primarily on $\lambda=4.5$ and $\lambda=6$. The wind turbine model had a nacelle and tower to imitate a real wind turbine as closely as possible within laboratory scale constraints. We report four main results: (a) the spatial region over which different frequencies are dominant in the near field varies drastically with $\lambda$. We introduce a length scale termed as the `convective pitch' which depends on $\lambda$ (and hence the spatial unfolding of the tip vortices) and show that it could be a better length scale than turbine diameter (D) to demarcate the near wake. (b) the free stream turbulence intensity for the experiments was negligible ($<1\%$), however, we still observed wake meandering. The Strouhal number of wake meandering decreased with tip speed ratio ($\lambda$). (c) the wake meandering frequency is found to be present even very close to the nacelle, upholding the notion that the nacelle is important to `seed' wake meandering. (d) the tower acts as an important source of asymmetry, resulting in a downward bending of the mean wake centerline, and increased turbulence and mixing in the lower plane.

\section{Experimental method}\label{Section_model}

\begin{figure}
  \centerline{
  \includegraphics[clip = true, trim = 0 0 0 0 ,width= 1.1\textwidth]{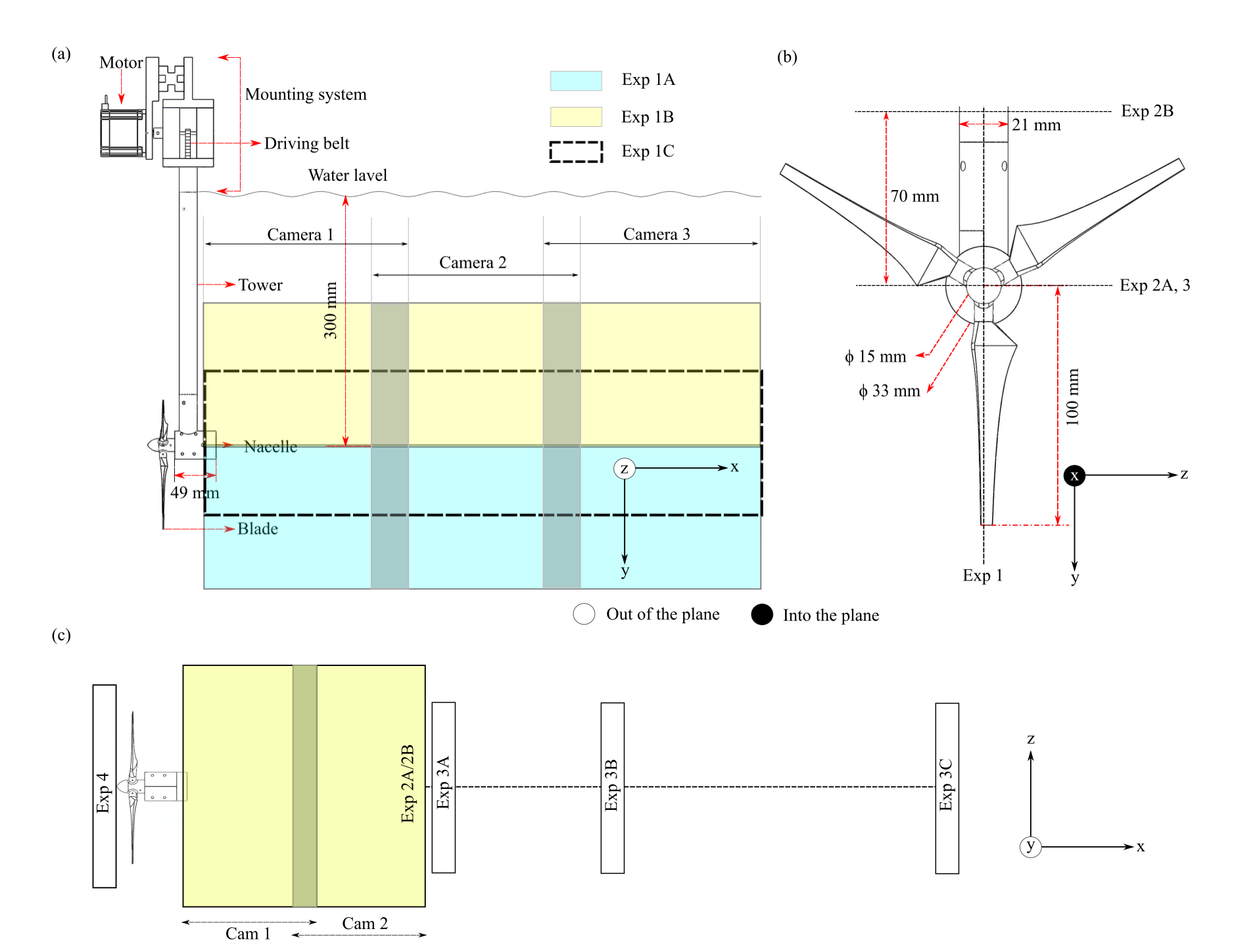}  }
 \caption{A schematic of the wind turbine model and the field of views used in the different experiments. The streamwise distance $x$ is measured from the rotor plane and transverse distances $z$ and $y$ are measured from the nacelle centerline. Experiment 1 (A-C) focused on the plane aligned with the tower's axis and streamwise direction, $\textit{i.e.}$ $xy$ plane (sub figure \ref{fig:sch}(a)), while experiments 2-4 focused on $xz$ planes at different $y$ offsets (sub figure \ref{fig:sch}(b-c)).}
\label{fig:sch}
\end{figure}

A series of particle image velocimetry (PIV) experiments were performed on a small scale wind turbine model in the hydrodynamics flume in the Department of Aeronautics at Imperial College London. At the operating water depth, the flume has a cross section of 60$\times$60 cm$^2$. A schematic of the model wind turbine is shown in figure \ref{fig:sch} which was designed to mimic the design of an actual wind turbine to the closest extent possible while satisfying several experimental constraints. The diameter of the model was restricted to 20 cm to keep the blockage low ( 8.7$\%$  based on turbine diameter which is comparable to blockages encountered in previous experimental studies \citep{sherry2013interaction, miller2019horizontal}). Note that the actual blockage is smaller as the rotor can be considered as a porous body. The freestream velocity ($U_{\infty}$) was kept constant at 0.2 m/s which led to a global Reynolds number (based on turbine diameter) of 40,000 which is several orders of magnitude smaller than the Reynolds number actual wind turbines operate at. The turbine was specifically designed to operate at low Reynolds number which is discussed in the next paragraph. The nacelle was represented as a cylindrical body of diameter 3.3 cm and length 4.9 cm. A hollow cylindrical tower of outer diameter 2.1 cm was attached to the nacelle. The tower was attached at the top to a mounting frame and the wind turbine model was hung in an inverted fashion in the flume. A stepper motor RS 829-3512 was used along with a drive and signal generator to rotate the turbine at a prescribed RPM. The motor, along with the speed controlling electronics, were located outside the flume and the torque from the motor was transmitted to the turbine shaft via a belt and pulley mechanism that was housed inside the hollow tower which restricted any further reduction in the tower's diameter.

Standard wind turbines operate at a high chord based Reynolds number $\sim 10^6$ (See fig:1 of \citep{lissaman1983low}) and at such Re standard airfoils operate at high maximum lift to drag ratio O(100) (See table 1 of \citep{sunada1997airfoil}) which is impossible to achieve in a small scale model (for which the chord based $Re$ is barely $\sim 9000$ here) unless a pressurised facility is used to tailor the density of the incoming flow \citep{miller2019horizontal}. Accordingly, there is inherently a Reynolds number mismatch of the order of 100. At $Re$ $\sim 10^4$, thin flat plate airfoils perform better than standard thicker smooth airfoils  \citep{mcmasters1980low, winslow2018basic, sunada1997airfoil, hancock2014wind}. Thus, a flat plate airfoil with thickness ratio $5\%$ and camber ratio $5\%$ was initially chosen for the blade which gives the best performance at low Reynolds number ($Re \sim 4 \times 10^3$) \citep{sunada1997airfoil}. However the blades were found to be incapable of sustaining the structural loads under the present experimental conditions and hence the thickness ratio was increased to $10\%$ for more structural rigidity. The operating angle of attack was selected to be $5^o$. The chord and twist distribution of the blades were similar to that used by \citet{hancock2014wind}. Near the root section, the blade was linearly interpolated to a circular section which was fixed to the hub. Experiments were conducted at tip speed ratios in the range $4.5\leq \lambda\leq 6$.

\begin{table}
  \begin{center}
\def~{\hphantom{0}}
  \begin{tabular}{lcccccccc}
      Exp  & $U_{\infty}(m/s)$ & $\lambda$ & FOV & Plane & $\delta x$ & $f_{aq}(Hz)$ & $\delta t(s)$ & $T(s)$ \\[3pt]
       
    1A   & 0.2 & 4.5, 6 &  \hspace{0.2 cm}\makecell{$0.25D<x<3.45D$,  \\ $0<y<0.85D$} \hspace{0.2 cm} & $z=0$ & $0.0088D$ & 100 & 0.01 & 54.75\\
       
    1B & 0.2 & 4.5, 6 & \hspace{0.2 cm}\makecell{$0.25D<x<3.45D$, \\ $-0.85D<y<0$}\hspace{0.2 cm} & $z=0$  & $0.0088D$ & 100 & 0.01 & 54.75\\ 

     1C & 0.2 & 4.5, 6 & \hspace{0.2 cm}\makecell{$0.25D<x<3.45D$, \\ $-0.42D<y<0.43D$}\hspace{0.2 cm} & {$z=0$}  & $0.0088D$ & 100 & 0.01 & 54.55\\

    2A & 0.2 & 4.5, 6 & \hspace{0.2 cm}\makecell{$0.29D<x<1.95D$, \\ $-0.73D<y<0.69D$}\hspace{0.2 cm} & {$y=0$}  & $0.0092D$ & 100 & 0.01 & 54.55\\

    2B & 0.2 & 4.5, 6 & \hspace{0.2 cm}\makecell{$0.29D<x<1.95D$, \\ $-0.73D<y<0.69D$}\hspace{0.2 cm} & {$y=-0.35D$}  & $0.0092D$ & 100 & 0.01 & 54.55\\

    3A & 0.2 & 4.5-6 & \makecell{$x\in(2D\pm 0.062D)$, \\ $-0.6D<y<0.6D)$} & $y=0$ & $0.0082D$ & 10 & 0.01 & 900\\

     3B & 0.2 & 4.5-6 & \makecell{$x\in(3D\pm 0.062D)$, \\ $-0.6D<y<0.6D)$} & $y=0$ & $0.0082D$ & 10 & 0.01 & 900\\

    3C & 0.2 & 4.5-6 & \makecell{$x\in(5D\pm 0.062D)$, \\ $-0.6D<y<0.6D)$} & $y=0$ & $0.0082D$ & 10 & 0.01 & 900\\

   4 & 0.2 & 4.5-6 & \makecell{$x\in(-1.2D\pm 0.062D)$, \\ $-0.7D<y<0.7D)$} & $y=0$ & $0.0082D$ & 10 & 0.01 & 120\\
     
  \end{tabular}
  \caption{parameters associated with different experiments. Here $\delta x$ represents the spatial resolution of the experiments. $f_{aq}$, $\delta t$, and $T$ are the acquisition frequency, time between successive laser pulses and total time of data acqiusition respectively.  }
  \label{tab:kd}
  \end{center}
\end{table}

Four different experimental campaigns, named campaigns 1 - 4, were conducted to capture different regions and properties of the flow. The details of the experiments can be found in table \ref{tab:kd} and fig. \ref{fig:sch}. In campaign 1, three Phantom v641 cameras were used to obtain a stitched field of view spanning $0.25 \leq x/D \leq 3.45$ in the streamwise direction in the $xy$ plane (see fig. \ref{fig:sch}), the distance being measured from the rotor plane. Here, in the plane of symmetry ($z=0$), three different experiments were conducted at each tip speed ratio considered. The first experiment in campaign 1 (henceforth referred to as $1A$) focused on the region $0\leq y/D \leq 0.85$, measured from the symmetry line. Similarly experiment $1B$ focused on the region $-0.85 \leq y/D \leq 0$ which contains the wake region of the tower. The 3rd experiment ($1C$) covered the central region, $-0.42 \leq y/D \leq 0.43$. In experimental campaign 2, two Phantom v641 cameras were used simultaneously to capture the near wake ($0.29 \leq x/D \leq 1.95$) in the $xz$ plane. Experiment $2A$ focused on the symmetry plane ($y=0$), while experiment $2B$ focused on an offset plane ($y=-0.35D$) such that some influence from the tower-wake is captured. All the parameters associated with different experiments are tabulated in table \ref{tab:kd}. For experiments 1 and 2, each camera captured images (of dimension $2560 px\times 1600 px$) in cinematographic mode at an acquisition frequency ($f_{aq}$) of $100$ $Hz$ which was found to be adequate to resolve all scales of dynamic importance for the present study. Data was obtained for a total time ($T$) of 54.75$s$ and 54.55$s$ for experiments 1 and 2 respectively. 

The objective of campaign 3 was to obtain the wake meandering frequency accurately. Only one camera was used for each experiment and the field of view was shrunk into a thin strip of dimension $2560 px \times 256 px$ (see fig. \ref{fig:sch}(c)) which facilitated obtaining large time series of data ($900 s$, which is close to 180 wake meandering cycles) within the memory constraints of the camera. The tip speed ratio was slowly varied from 4.5 to 6 in increments of 0.1. The acquisition frequency was reduced to $10Hz$ and the time between successive laser pulses ($\delta t$) was kept fixed at $0.01s$ for all experiments. The field of views of the experiments were centred at $x=2D$, $3D$ and $5D$ as shown in fig. \ref{fig:sch}. Finally experiment 4 used a similar strip field of view just upstream of the rotor (see details in table \ref{tab:kd}) which is used to obtain the approach velocity at different tip speed ratios. The images acquired were processed in PIVlab \citep{thielicke2014pivlab}. The adaptive cross correlation algorithm in PIVlab used a multi-pass, fast Fourier transform to determine the average particle displacement. An initial interrogation area (IA) of $64\times64$ pixels was reduced in $3$ passes with a final IA size of $16\times16$ pixels with $50$\% overlap in the $x$ and $y$ directions. The spatial resolution ($\delta x$) of the experiments were close and were around $0.0082D-0.0092D$ (see table \ref{tab:kd}). The smallest scales of dynamic importance in the near field, \textit{i.e.} the tip vortex cores were found to span $\approx$ $6 \delta x$. Since the interrogation windows had a $50\%$ overlap, \NB{meaning that adjacent vectors in the velocity fields were spaced $\delta x/2$ apart}, the tip vortex cores spanned $\approx$ 12 PIV vectors. Accordingly, all scales of dynamic importance are believed to be resolved adequately.

\section{Results and discussion}

\begin{figure}
  \centerline{
  \includegraphics[clip = true, trim = 0 0 50 0 ,width= \textwidth]{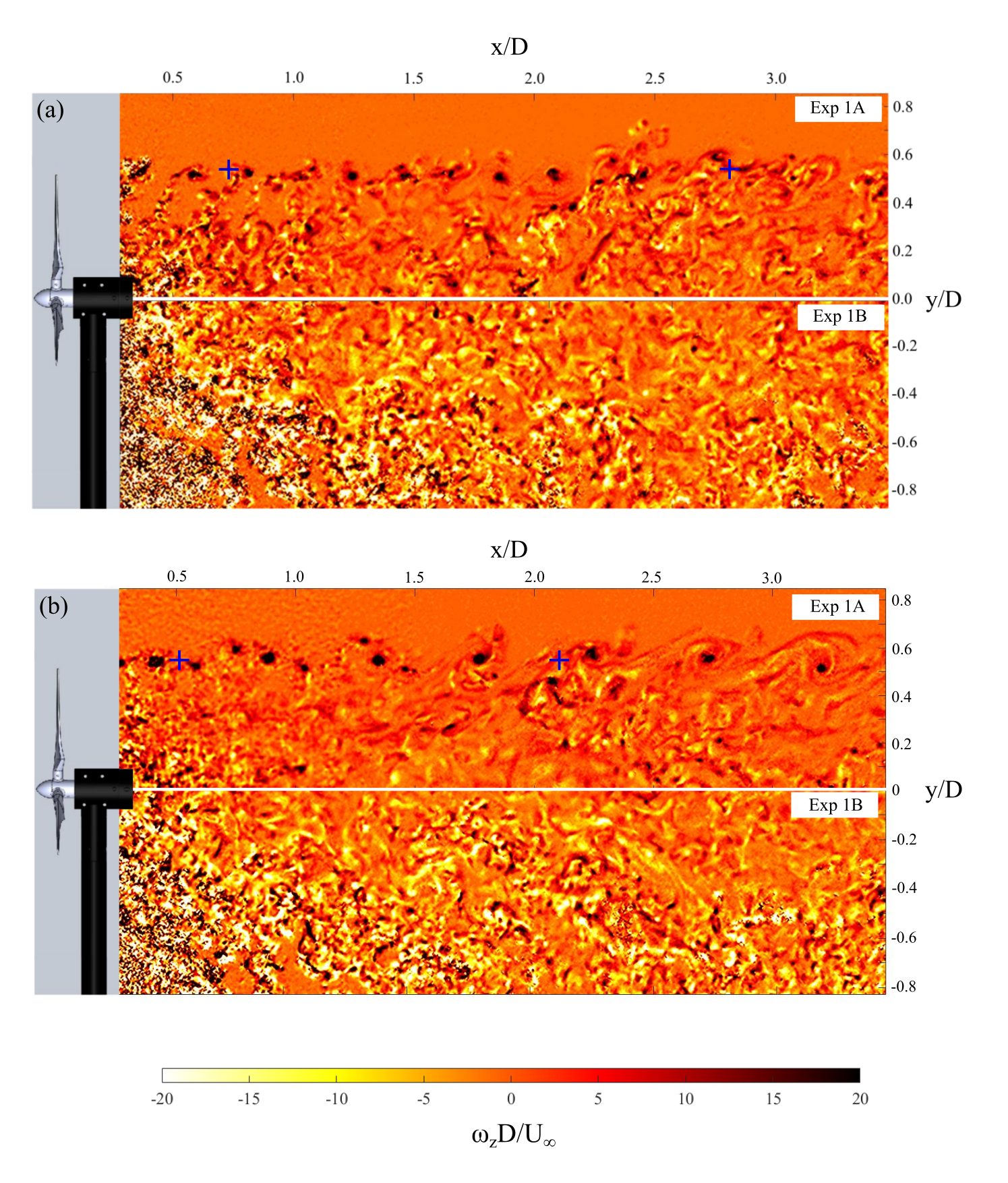}  }
 \caption{Instantaneous vorticity fields of a wind turbine wake for (a) $\lambda=4.5$ and (b) $\lambda=6$ in the $xy$ plane. Field of views from experiment $1A$ and experiment $1B$ (at different time instants) are stitched together for a visual representation of the entire wake. }
\label{fig:inst}
\end{figure}

\subsection{Instantaneous vorticity field}

Fig. \ref{fig:inst}(a-b) show instantaneous vorticity fields at $\lambda = 4.5$ and $\lambda=6$ at the $xy(z=0)$ plane. Note that the field of views from experiments $1A$ and $1B$ are stitched together for visual representation but they were not acquired concurrently. The flow fields in fig. \ref{fig:inst} are inherently complex and contain several length/time scales (also see supplementary video 1 and 2). In the top plane (experiment $1A$), the array of the tip vortices can be seen which acts as a boundary between the wind turbine wake and the free stream. For $\lambda=4.5$, the vortices shed from the 3 blades start interacting at a streamwise distance of $x/D \approx 2.25$ and initiate the merging process. Until $x/D \approx 2.25$, the wake boundary remains nearly horizontal or in other words, in the presence of the tip vortices, the wake does not expand in the very near field. It supports the observation of \citep{medici2005experimental, lignarolo2015tip} who noted that the tip vortices in the near field act as a shield to prevent mixing with the outer fluid. Beyond $x/D \approx 2.25$ the interaction and merging of the tip vortices aid wake expansion and wake recovery. For $\lambda=6$, the tip vortices are stronger and are more closely spaced (due to higher rotational speed), which results in an earlier interaction and merging (see supplementary video 2). In the lower plane (experiment $1B$), the vorticity field looks drastically different from the top plane. The presence of the tower causes an early breakdown of the tip vortices. Note that the vorticity levels in the lower plane are significantly enhanced compared to the upper plane and the tower acts as an important source for this asymmetry in the flow.

\begin{figure}
  \centerline{
  \includegraphics[clip = true, trim = 0 50 0 0 ,width= \textwidth]{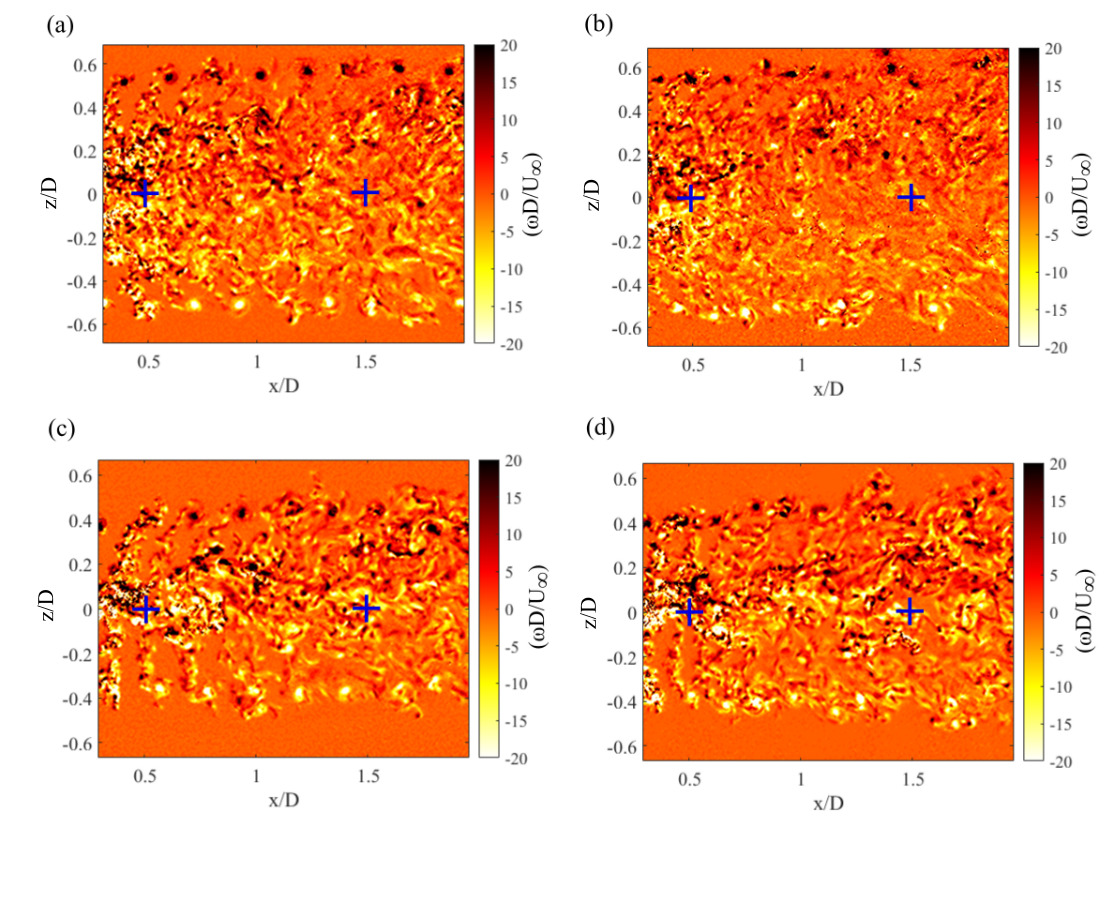}  }
 \caption{Instantaneous vorticity field of a wind turbine wake in $xz(y=0)$ plane for (a) $\lambda=4.5$ and (b) $\lambda=6$. Subfigures (c) and (d) show the vorticity field at an offset plane ($y=-0.35D$) for $\lambda=$ 4.5 and 6 respectively.}
\label{fig:inst2}
\end{figure}

Fig. \ref{fig:inst2} shows the instantaneous vorticity field in the $xz$ plane at different $y$ offsets at $\lambda=4.5$ and 6 (Exp 2). At the symmetry plane (experiment $2A$) the flow field looks similar to that obtained from experiment $1A$ (see supplementary videos 3 and 4) and the wakes are mostly symmetric about the nacelle centerline ($z=0$) (see fig. \ref{fig:inst2}(a, b)). Fig. \ref{fig:inst2}(c, d) show the vorticity field at an offset (experiment $2B$) from the nacelle. The central region of the wake shows an oscillatory behaviour (see supplementary videos 5 and 6) which was not pronounced at the symmetry plane in experiment $2A$. It is evident that this oscillation results from the vortex shedding of the tower which interacts with the tip vortices. This interaction causes an earlier breakdown of the tip vortices, promoting turbulence and mixing in the lower plane seen in fig. \ref{fig:inst}. It is worthwhile to note that the flow field shown in fig. \ref{fig:inst} or \ref{fig:inst2} is similar to that observed in an actual turbine (see fig. 7 of \citep{abraham2019effect}). \citet{abraham2019effect} utilised natural snowfall to visualise the wake of an utility scale turbine in the symmetry plane ($xy$ plane according to present nomenclature) and showed that the flow structures in the lower plane were significantly more chaotic and distorted due to the presence of the tower. Hence, we believe the present experiments replicate the wake of an actual turbine well in spite of the inherent Reynolds number difference.

\begin{figure}
  \centerline{
  \includegraphics[clip = true, trim = 0 100 200 50 ,width= 1.2\textwidth]{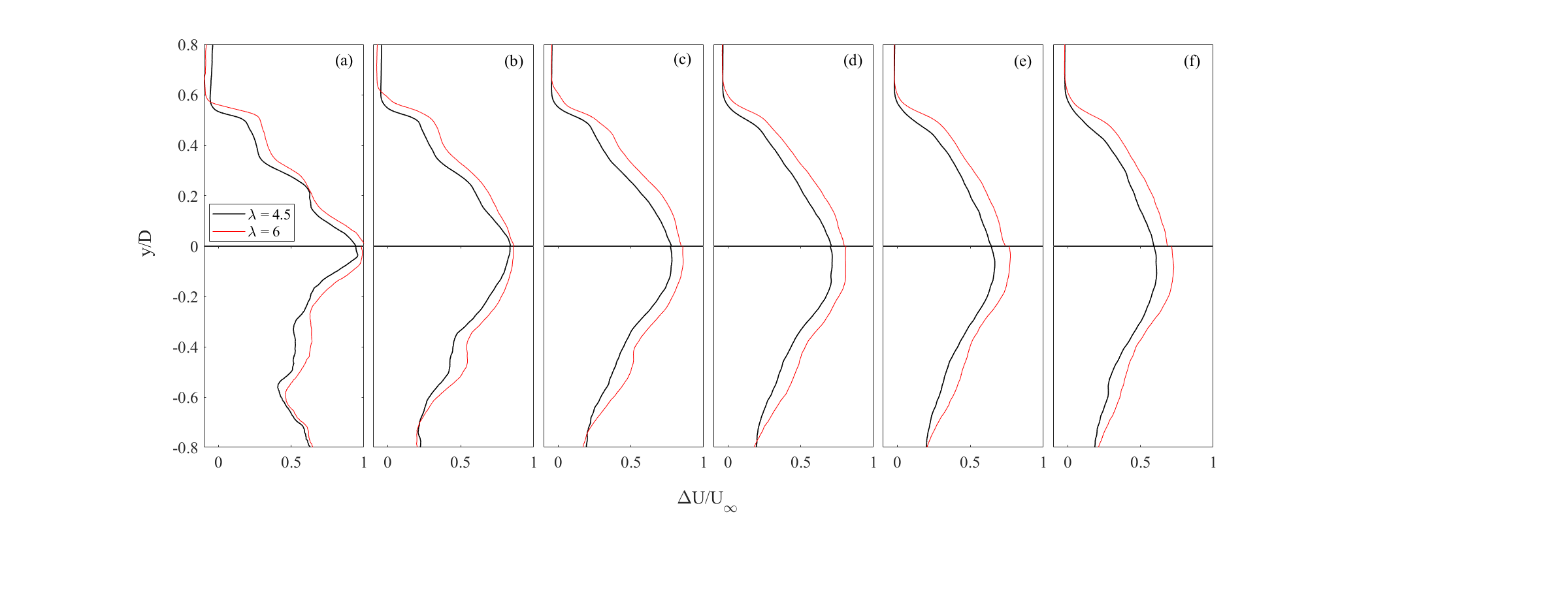}  }
 \caption{ Mean velocity deficit at (a) $x/D = 0.5$, (b) $x/D = 1.0$, (c) $x/D = 1.5$, (d) $x/D = 2.0$, (e) $x/D = 2.5$, and (f) $x/D = 3.0$ for $\lambda = 4.5 $ and $6$}
\label{fig:def}
\end{figure}

\subsection{Mean velocity field}
Fig. \ref{fig:def} shows the mean velocity deficit ($\Delta U/U_{\infty}$, where $\Delta U = U_{\infty} - U$) profiles at different streamwise distances for the two different tip speed ratios. The plots are only shown in the $xy$ plane (experiment 1). Close to the turbine, in the upper plane, 3 distinct inflection points are observed (fig. \ref{fig:def}(a)). We know that the mean velocity/velocity deficit profiles in the wake of a single scale bluff body contains only one inflection point on each side of the wake. The existence of multiple inflection points in the wake of the wind turbine essentially shows the multiscale nature of a wind turbine wake. \citet{baj2017interscale} observed mean velocity profiles of similar nature in the near wake of a multiscale array of prisms. Note that the qualitative nature of the wake deficit profiles are similar for both tip speed ratios considered. In the upper plane, the inflection point near the centreline occurs at $y/D \approx 0.1$, which is close to the surface of the nacelle. Hence it corresponds to the wake of the nacelle, where the velocity deficit is maximum. The second inflection point occurs at $y/D \approx 0.3$ which is close to the root region of the blade and corresponds to the blade wake. The third inflection point occurs at $y/D \approx 0.55$ and corresponds to the tip vortices. The first inflection point is the least persistent (until $x/D \approx 1$) indicating a small spatial extent of the nacelle wake. The second inflection point persists until $\approx$ 1.5 diameters, while the 3rd inflection point persists far beyond that, which can be expected as the tip vortices are more spatially persistent than the vortices shed from the root region of the blade, as can also be seen from fig. \ref{fig:inst}. Important observations can be made if we compare the mean velocity deficit profiles with standard wake models used in industry. Standard models like Jensen \citep{jensen1983note} or Frandsen \citep{frandsen2006analytical} assume that the velocity deficit has a symmetric top hat shape which is not true even at $x/D \approx 3$. The model proposed by \citet{bastankhah2014new} assumes a Gaussian distribution of the velocity deficit profile, which is by far a better assumption to make but still has limitations. As can be seen from figure \ref{fig:def}, even at $x/D = 3$, the wake deficit profile is not exactly Gaussian. Sharp gradients exist near the wake edge as the tip vortices are still coherent. Additionally, the location where velocity deficit is maximum drifts from the geometric centreline ($y=0$) towards $y<0$. This asymmetry is caused by the presence of the tower which is not accounted for in the aforementioned models in use. Indeed, the tower is the most important source of top/bottom asymmetry in the wake which leads to higher velocity deficit in the lower plane that persists far downstream.


\begin{figure}
  \centerline{
  \includegraphics[clip = true, trim = 0 0 0 0 ,width= 1.05\textwidth]{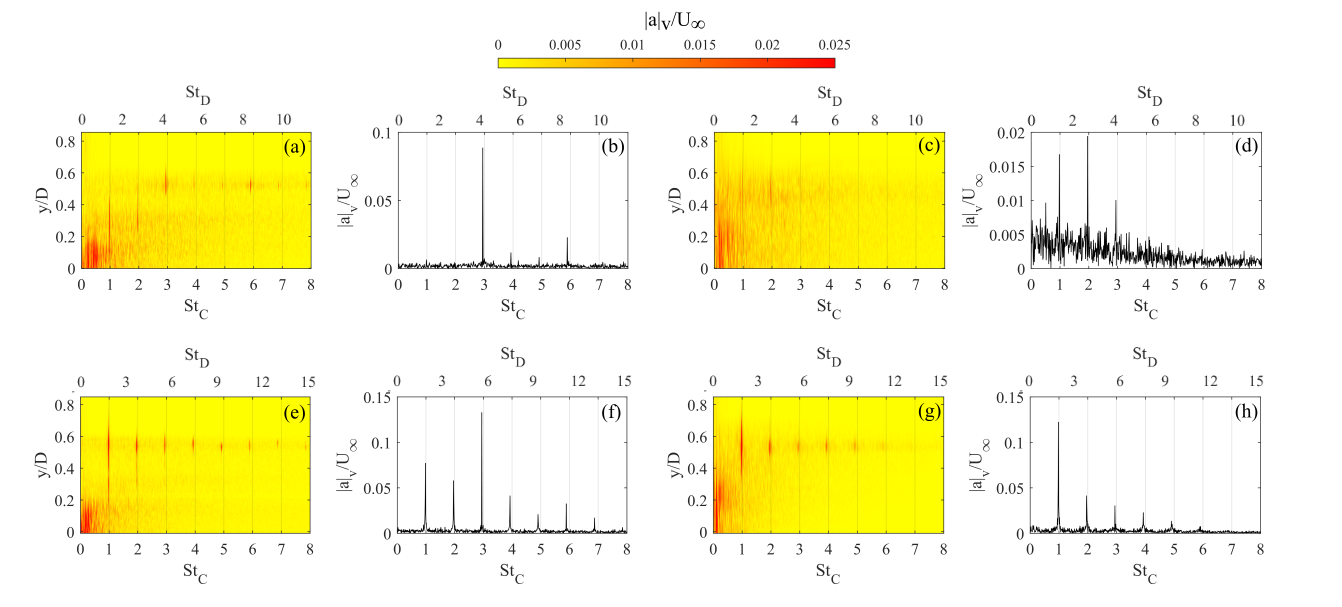}  }
 \caption{Transverse velocity spectra obtained at $x/L_c=1$ for (a) $\lambda=4.5$ and (e) $\lambda=6$. Subfigure (b) and (f) show the same at $x/L_c=1$, $y/D=0.55$ for the two tip speed ratios. Similarly, subfigure (c) and (g) show the spectra at $x/L_c=4$ for $\lambda = 4.5$ and $\lambda = 6$ respectively. The corresponding spectra at $x/L_c=4$, $y/D=0.55$ are shown in (d) and (h). Strouhal numbers $St_C$ and $St_D$ are defined based on $L_c$ and $D$ as the length scale and free stream velocity as the velocity scale.}
\label{fig:fft_turb_1}
\end{figure}

\begin{figure}
  \centerline{
  \includegraphics[clip = true, trim = 0 0 0 0 ,width= 0.9\textwidth]{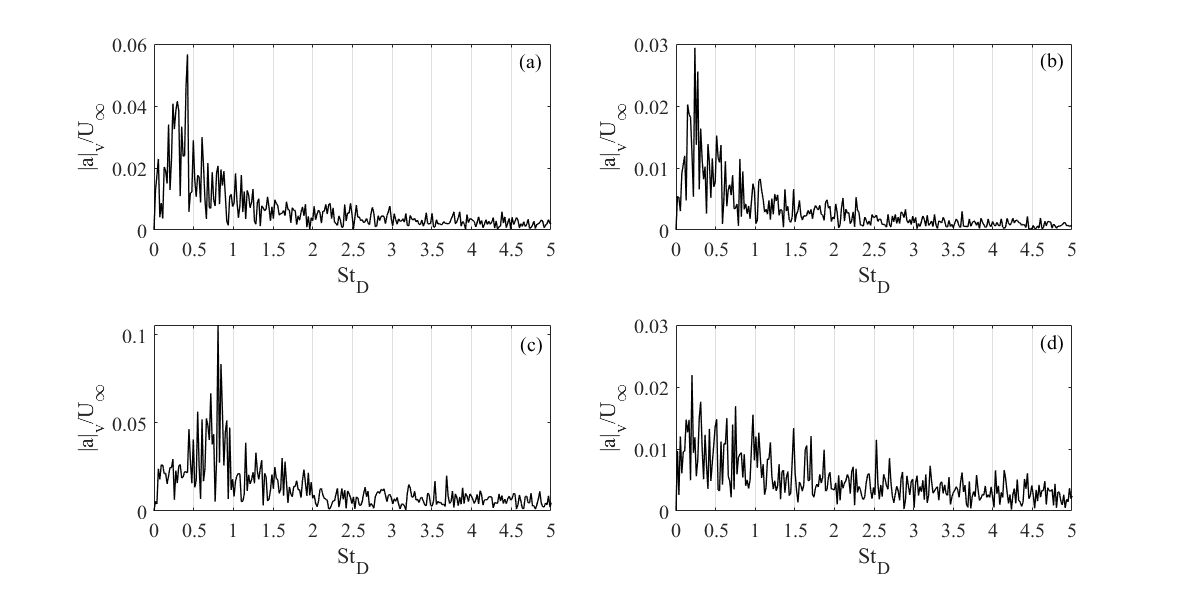}  }
 \caption{Transverse velocity spectra obtained at (a) $x/D=0.5$, $z/D=0$ and (b) $x/D=1.5$, $z/D=0$ at the plane $y=0$ (experiment $2A$). The same is shown at an offset plane (experiment $2B$) at (c) $x/D=0.5$, $z/D=0$, and (d) $x/D=1.5$, $z/D=0$. All the spectra are shown for for $\lambda = 6$ only.}
\label{fig:fft_turb_2}
\end{figure}

\subsection{Important frequencies}

    The temporal fluctuation in the wake of a wind turbine is important as that determines the nature of the fluctuating loads induced on downstream turbines exposed to the wake of the upstream machine. Due to the inherent multiscale nature of the wake, multiple frequencies can be expected to characterise the wake dynamics. Let us first identify the important frequencies towards the outer edge of the wake for different tip speed ratios. For that, we introduce a length scale, which we term as \emph{Convective pitch, $L_c$} and define it as $L_c=\pi D/\lambda$. This length scale can physically be interpreted as the distance travelled by a fluid element in the time taken for the rotor to complete one full rotation. For the present configuration $L_c = 0.70D$ and $0.52D$ for $\lambda=$ 4.5 and 6 respectively. We evaluate fast Fourier transforms at selected points (shown by \textbf{\textcolor{blue}{+}} in fig. \ref{fig:inst}) based on the fluctuating transverse velocity component and show them in fig. \ref{fig:fft_turb_1}. The Strouhal number, $St_C$ is calculated based on freestream velocity ($0.2$m/s) and $L_c$.  Figs. \ref{fig:fft_turb_1}(a) and \ref{fig:fft_turb_1}(e) show the transverse velocity spectra at $x/L_c = 1$ for $\lambda = 4.5$ and $\lambda=6$ respectively. Apart from a low frequency region near the nacelle ($y\approx 0$), a number of distinct frequencies are observed which match well with integer multiples of $St_C$. These frequencies and their relative strengths can be more clearly observed from fig. \ref{fig:fft_turb_1}(b) and \ref{fig:fft_turb_1}(f) (for $\lambda = 4.5$ and $\lambda=6$ respectively) which show the frequency spectra at $x/L_c=1$ and $y/D=0.55$, \textit{i.e.} near the outer wake. Note that, the spectra look qualitatively similar for both tip speed ratios. Here, the dominant frequency is $St_C \approx 3$, which corresponds to the blade passing frequency (henceforth denoted as $3f_r$, where $f_r$ is the rotor frequency). This similarity holds up the possibility to demarcate the very near field of the turbine wake (where the influence of the blade passing frequency is significant) based on the convective pitch, $L_c$, which we discuss further in section \ref{freq_map}. Although the dominant frequency is the same for both $\lambda$ at this location, there are subtle differences in the spectra. Especially, for $\lambda=6$, the rotor frequency $f_r$ ($St_C \approx 1$) as well as other harmonics of blade/rotor frequencies are more pronounced compared to $\lambda=4.5$.

    \begin{figure}
  \centerline{
  \includegraphics[clip = true, trim = 0 0 0 0 ,width= 0.95\textwidth]{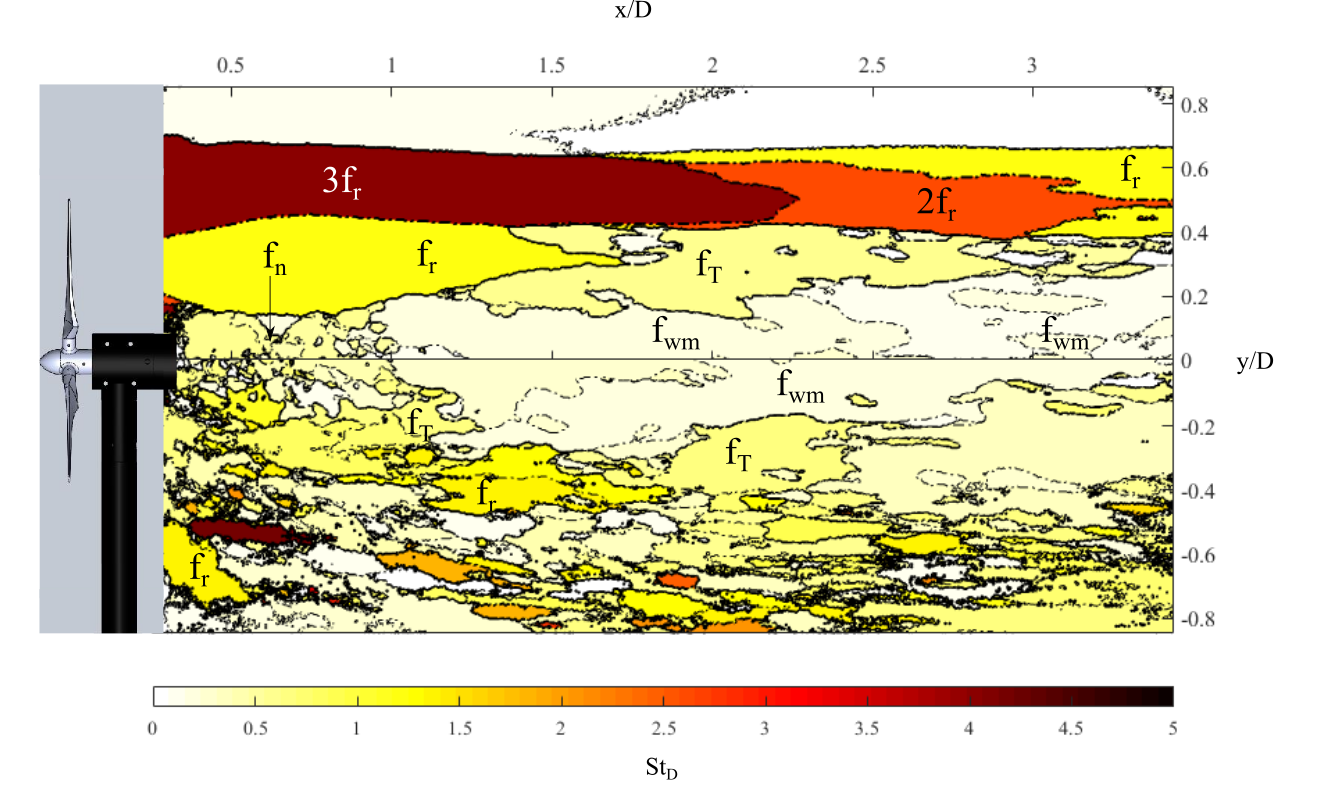}  }
 \caption{ Zones of dominant frequencies for $\lambda = 4.5$ }
\label{fig:spat_4pt5}
\end{figure}

    The difference between the two tip speed ratios becomes more significant in the far field as can be seen from figs. \ref{fig:fft_turb_1}(c) and \ref{fig:fft_turb_1}(g) which show the transverse velocity spectra at $x/L_c = 4$. The tip vortices are much stronger for $\lambda=6$ here. The corresponding spectra at $x/L_c=4$ and $y/D=0.55$ are shown in fig. \ref{fig:fft_turb_1}(d) and \ref{fig:fft_turb_1}(h). It can be seen that the blade passing frequency ($3f_r$) is no longer dominant. For $\lambda=4.5$, the dominant frequency is $2f_r$, while $f_r$ is comparable to $2f_r$. For $\lambda=6$ however, $f_r$ is by far the dominant frequency at this location, which implies that, the merging process for $\lambda=6$ is fundamentally different from $\lambda=4.5$. Different modes of tip vortex merging have been reported in experiments \citep{sherry2013interaction, felli2011mechanisms} and theory \citep{widnall1972stability}. The merging process is known to be primarily driven by mutual inductance of the helical tip vortices and it has been shown to depend on vortex strength, vortex core size and pitch of the vortices \citep{widnall1972stability}. \citet{felli2011mechanisms, sherry2013interaction} argued that the merging of the tip vortices is a two-step process, where two vortex filaments get entangled first and thereafter merge with the third filament further downstream leading to a single vortical structure. The dominance of $2f_r$ in the far field for $\lambda=4.5$ is believed to be the result of such a two-step merging process. However, this is not very evident from supplementary video 1, as the vortex cores of the tip vortices for $\lambda=4.5$ are weaker compared to $\lambda=6$ and they quickly get diffused. For $\lambda=6$, the vortex cores are much stronger and their separation is shorter, as a result of which there is a stronger and earlier interaction. From supplementary video 2 ($\lambda=6$) it can be noted that two tip vortices start revolving around the third tip vortex and eventually merge into a single vortical structure hinting at a single-step merging. Our data leads us to believe that the merging process may or may not be a multistage process depending on the tip speed ratio, other geometric factors such as specific design of the turbine may also influence how the tip vortices ultimately merge.

\begin{figure}
  \centerline{
  \includegraphics[clip = true, trim = 0 0 0 0 ,width= 0.9\textwidth]{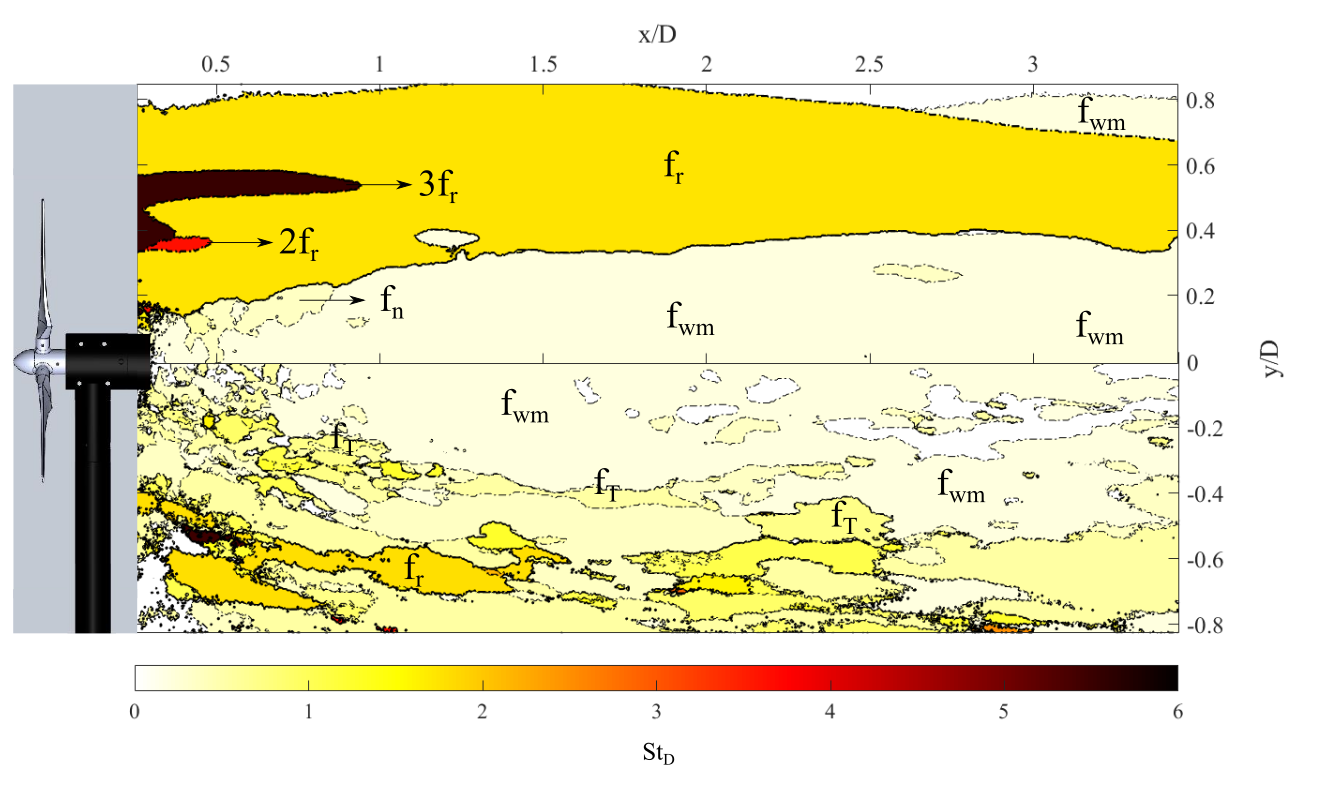}  }
 \caption{ Zones of dominant frequencies for $\lambda = 6$.}
\label{fig:spat_6}
\end{figure}

    Let us now look at the frequencies near the central region of the wake. For that we perform fast Fourier transforms at selected points from experiment 2, which are shown by the \textbf{\textcolor{blue}{+}} sign in fig. \ref{fig:inst2}. The results are shown in fig. \ref{fig:fft_turb_2} and are only for $\lambda=6$, as the corresponding results for $\lambda=4.5$ were similar. Hence, unlike fig. \ref{fig:fft_turb_1}, the Strouhal number is calculated only based on diameter $D$. Figs. \ref{fig:fft_turb_2}(a) and \ref{fig:fft_turb_2}(b) are obtained from experiment $2A$ ($y=0$ plane) at $x/D=0.5,z/D=0$ and $x/D=1.5,z/D=0$ respectively. Near the nacelle (fig. \ref{fig:fft_turb_2}(a)) the dominant frequency is at $St_D=0.42$. Note that the nacelle diameter for this study is close to 6 times smaller than the rotor diameter. Hence, if we use the nacelle diameter to scale the frequency instead, the Strouhal number comes to around $0.069$. \citet{abraham2019effect} reported a similar nacelle shedding frequency $St \approx 0.06$ for a utility scale turbine. Accordingly, we believe this frequency is related to vortex shedding from the nacelle and henceforth denote it as $f_n$. Interestingly, away from the nacelle, at $x/D=1.5,z/D=0$ (fig. \ref{fig:fft_turb_2}(b)), an even lower dominant frequency is observed at $St_D \approx 0.23$. This Strouhal number correlates well with the wake meandering frequencies reported in several previous studies \citep{okulov2014regular, chamorro2009wind} and is henceforth denoted as $f_{wm}$. The origin of $f_{wm}$ is discussed in more detail in section 4. Figures \ref{fig:fft_turb_2}(c)-(d) are obtained from experiment $2B$, \textit{i.e.} at an offset plane ($y=-0.35D$), away from the nacelle and downstream of the tower. In the near wake ($x/D=1.5,z/D=0$, fig. \ref{fig:fft_turb_2}(c)), a dominant frequency is found at $St_D \sim 0.8$. We argue that this frequency corresponds to the vortex shedding from the tower and denote it as $f_T$. Upon non-dimensionalisation based on tower diameter and $U_{\infty}$, the Strouhal number is around 0.084, which is significantly lower than the expected value of $St\approx 0.2$ for vortex shedding from a cylinder. Note that, the Strouhal number was calculated based on freestream velocity, however the flow ahead of the tower is strongly sheared due to the presence of the rotor. The incoming velocity just ahead of the tower was not measured in the experiments, but can be expected to be significantly lower than the freestream velocity, hence resulting in a local reduction in shedding frequency. Further downstream (fig. \ref{fig:fft_turb_2}(d)), $f_T$ is no longer prominent. The wake meandering frequency is still observed, but is weaker compared to the central plane ($y=0$), indicating a stronger influence of wake meandering in the central region of the wake.

\begin{figure}
  \centerline{
  \includegraphics[clip = true, trim = 20 0 0 0 ,width= 0.75\textwidth]{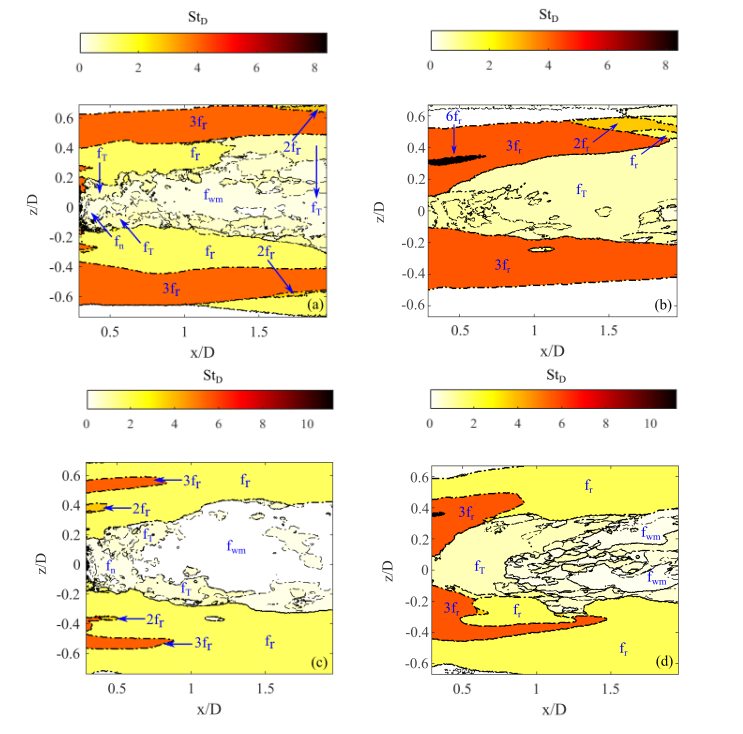}  }
 \caption{ Zones of dominant frequencies for $\lambda = 4.5$ at the plane (a) $y=0$ and (b) $y=-0.35D$. Subfigures (c) and (d) show the same for $\lambda=6$.}
\label{fig:spat_top}
\end{figure}

\subsubsection{Frequency maps}
\label{freq_map}
Having observed the presence of multiple frequencies in the near field, let us identify the zones in which a particular frequency is dominant. We define the dominant frequency to be that at which the transverse velocity spectrum attains its maximum value at a particular location. For instance, in fig. \ref{fig:fft_turb_1}(b) and \ref{fig:fft_turb_1}(f), \textit{i.e.} in the near field and in the tip region, the dominant frequency is $3f_r$, while relatively far from the rotor, near the central region (see fig. \ref{fig:fft_turb_2}(b)), it is the wake meandering frequency ($f_{wm}$), which is dominant. We obtain the dominant frequencies at all the spatial locations and create a frequency map demarcating zones where a particular frequency is dominant. The dominant frequency map for $\lambda =4.5$ in the $xy$ plane is shown in fig. \ref{fig:spat_4pt5}. On the top plane, large zones with distinct boundaries are observed where a particular frequency is dominant. Near the rotor in the tip region, $3f_r$ is dominant until around $x/D \approx 2.25$, beyond which $2f_r$ and $f_r$ become important. Around $x/D \approx 2.25$, the merging process of tip vortices initiates as was seen from fig. \ref{fig:inst}. Distinct root vortices (having a frequency of $3f_r$) have been reported in some studies \citep{sherry2013interaction} in the vicinity of the root region of the blade. We do observe a small region in the vicinity of the root region where $3f_r$ is dominant (see fig. \ref{fig:spat_4pt5}), however, root vortices were not pronounced for the present study, which can be seen from fig. \ref{fig:inst} (see also supplementary videos 1 and 2). We believe this has to do with the specific design of the blade in the root region that does not produce strong root vortices. Interestingly, near the root region of the blade, until around $x/D \approx 1.5$, there is a large region where $f_r$ is dominant. Note that, the porosity of the rotor disk is effectively low near the root region of the blade. As a result, there is a strong contribution from viscosity in driving the flow at the rotor frequency. Very close to the nacelle, there is a small region of nacelle shedding frequency ($f_n$). Further downstream, large regions are seen where the tower frequency ($f_T$) is dominant. The observation of $f_T$ in the $xy$ plane is particularly interesting as it indicates the inherent three-dimensionality of the vortex shedding from the tower. Several factors can lead to a three-dimensional vortex shedding from the tower such as the strongly sheared inflow condition \citep{silvestrini2004direct} caused by the rotor and finite-span effects of the tower near the nacelle \citep{williamson1996vortex}. It has been shown that the presence of shear or end effects can lead to oblique vortex shedding which can promote three-dimensionality in the wake \citep{silvestrini2004direct, williamson1996vortex}. Such oblique three-dimensional vortex shedding from the tower could be clearly observed in supplementary videos 1 and 2 (shown by a blue arrow). This obliqueness in the tower vortex shedding is also manifested in the form of oblique outward bursts from the centreline of the turbine wake. Such bursts can be repeatedly observed in supplementary videos 1 and 2 for $\lambda=4.5$ and $\lambda=6$ (shown by white arrows). 

\begin{figure}
  \centerline{
  \includegraphics[clip = true, trim = 0 0 0 0 ,width= 1.2\textwidth]{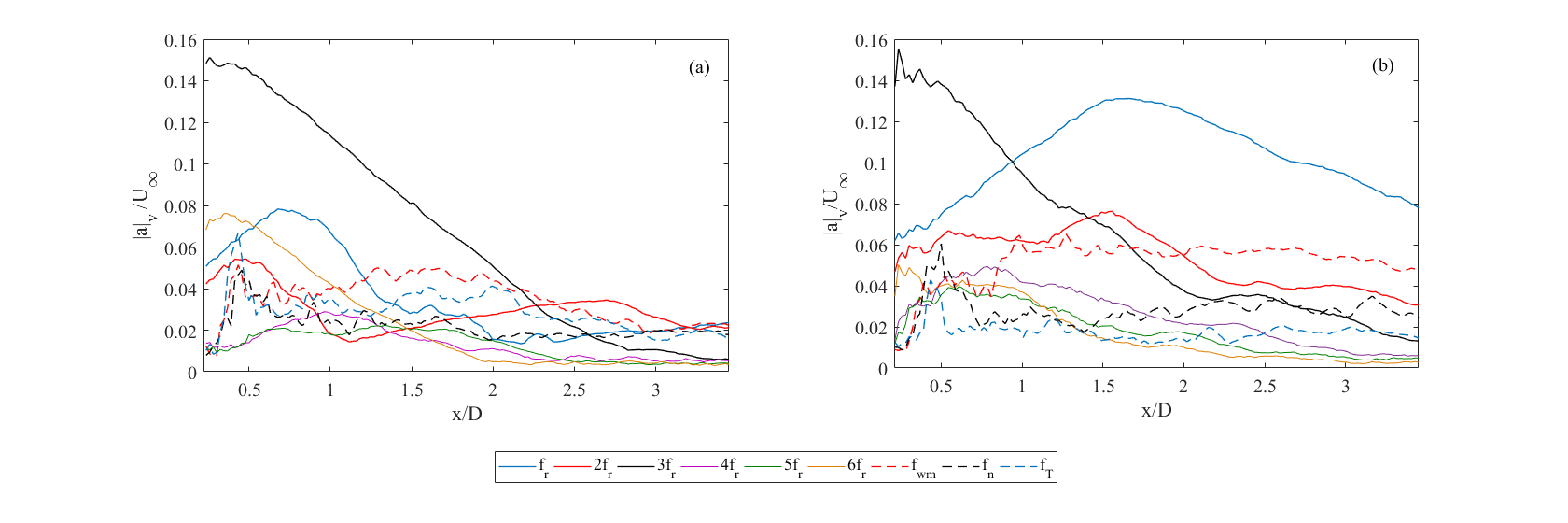}  }
 \caption{ The evolution of the `strength' of different frequencies with streamwise distance for (a) $\lambda=4.5$ (a) and (b) $\lambda=6$. The strength is calculated as the magnitude of the largest peak in the spectrum of transverse velocity at a particular streamwise location. }
\label{fig:evo_down}
\end{figure}

In the lower plane ($y<0$), the tip vortices break down earlier due to their interaction with the tower's vortex shedding. Patches of high frequency regions can be seen near the tip region in the lower plane, which are the remains of the blade passing or rotor frequencies. Apart from that, a number of other frequencies are also observed, making it look more broadband and turbulent like. The wake meandering frequency ($f_{wm}$) is found to be dominant only near the central region in the near wake. The entire wake looks like a shell of high frequency fluid surrounding the central region dominated by low frequency dynamics. 

The scenario remains qualitatively similar for $\lambda = 6$ as can be seen in fig. \ref{fig:spat_6}, that is the high frequency shell surrounds a low frequency zone. However, the extents of the dominant zones clearly look different. In the upper plane, the zones where $3f_r$ and $2f_r$ are dominant is quite small, and $f_r$ is dominant in a large portion of the upper plane. This is because at a higher $\lambda$, the tip vortices start to interact much earlier. Also, in the top plane the wake is much wider for $\lambda = 6$. In the lower plane, traces of the tip vortices are still observed but again there are several new frequencies present and it is more broadband. In the central region wake meandering is dominant again and the extent of the region where it is dominant grows with downstream distance. In fact, $f_{wm}$ remains dominant over a larger region at $\lambda=6$ compared to $\lambda=4.5$, which shows a possible dependence of wake meandering on the tip speed ratio. This dependence is further discussed in section \ref{wm}.

Next we investigate the frequency zones obtained from experiment 2. Figs. \ref{fig:spat_top}(a) and \ref{fig:spat_top}(c) correspond to $\lambda=4.5$ and $\lambda=6$ respectively at the central plane ($y=0$) and are similar to the upper plane in fig. \ref{fig:spat_4pt5} and fig. \ref{fig:spat_6}. This offers reassurance that the results of the experiments are reproducible. The wake meandering frequency remains dominant over a larger region for $\lambda=6$ compared to $\lambda=4.5$. The difference between the two tip speed ratios is more pronounced at the offset plane (experiment 2B). For $\lambda=4.5$, the tower frequency ($f_T$) dominates the central part of the wake (see fig. \ref{fig:spat_top}(b)). Contrastingly, for $\lambda=6$, $f_T$ has a sole dominance only before $x/D \approx 1$, beyond which the strengths of $f_T$ and $f_{wm}$ become comparable. This evidence indicates that the strength of wake meandering depends on the tip speed ratio. Note that for $\lambda=6$, the wake appears to be slightly deflected towards the side $z>0$ which could be caused by a slight mis-alignment of the rotor plane with the freestream direction in the experiments. The angle of deflection was measured to be $< 3^\circ$ and it does not alter the conclusions. 

\begin{figure}
  \centerline{
  \includegraphics[clip = true, trim = 0 0 0 0 ,width= 1.2\textwidth]{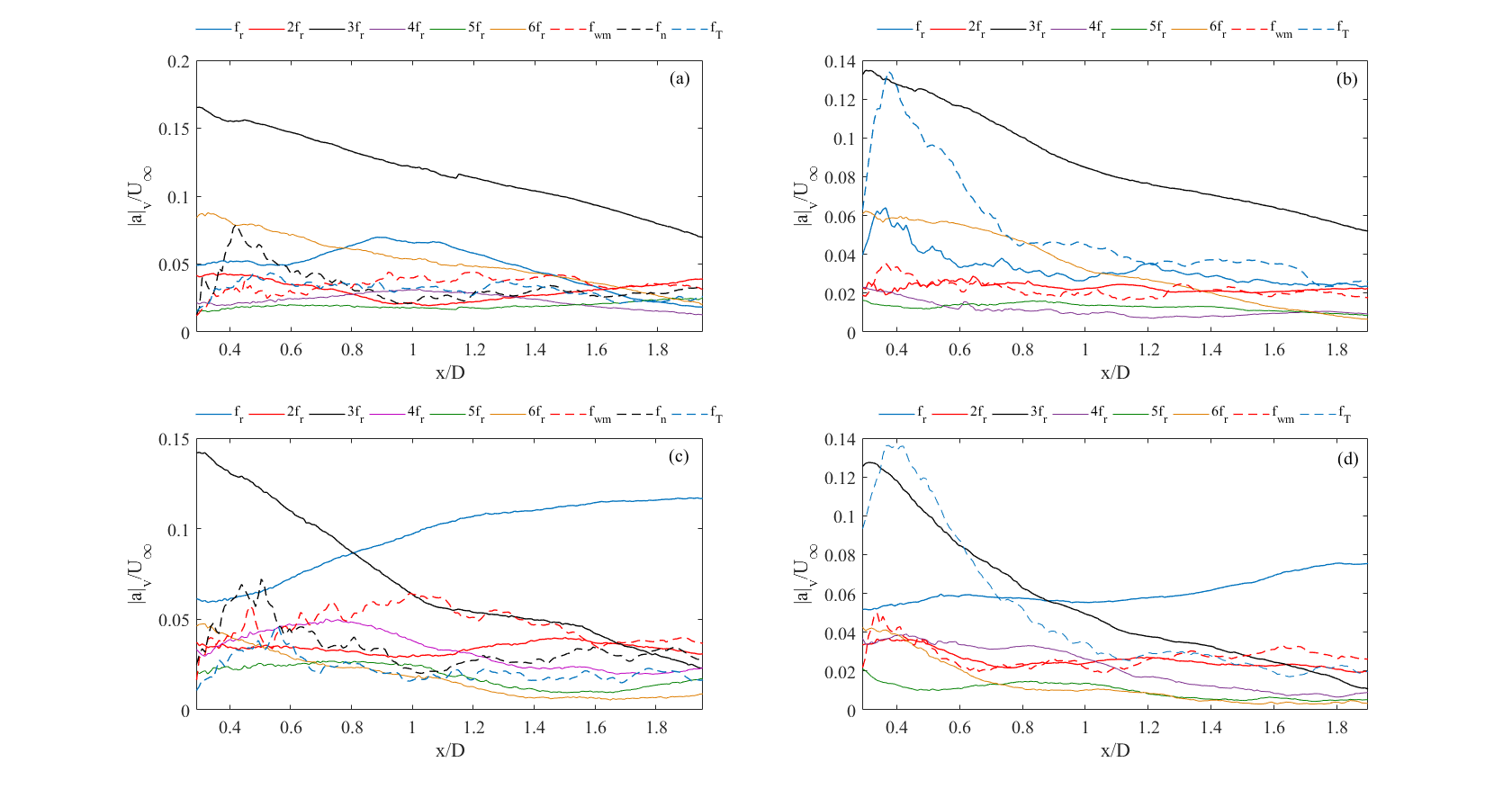}  }
 \caption{ Strength of different frequencies in the $xz (y=0)$ plane for (a) $\lambda = 4.5$ and (c) $\lambda = 6$ respectively. Sub figures (b) and (d) show the same at an offset plane $y = -0.35D$. }
\label{fig:evo_top}
\end{figure}

Let us cast a closer look at the streamwise evolution of the `strength' of different frequencies. The `strength' of a frequency can be measured locally by taking the maximum amplitude in the transverse velocity spectra of a particular frequency at a particular streamwise distance, which we denote as $|a|_{v}$. In fig. \ref{fig:evo_down}(a), $|a|_{v}$ is shown over streamwise distance in the upper plane (experiment $1A$) for $\lambda = 4.5$. In the near field, the blade passing frequency ($3f_r$) dominates over other frequencies and corresponds to the passage of tip vortices. At $x/D \approx 2.25$, $2f_r$ surpasses $3f_r$ and at $x/D \approx 3.2$, $f_r$ surpasses $2f_r$, which is indicative of the two-step merging process. The higher harmonics are comparatively damped and less important beyond $x/D \approx 1.5$. The wake meandering frequency is observed even very close to the nacelle and the local strength of the frequency does not change appreciably throughout the domain of investigation.

Fig. \ref{fig:evo_down}(b), shows $|a|_{v}$ for $\lambda = 6$. Similarly to $\lambda = 4.5$, the dominant frequency in the near field is $3f_r$. However, owing to the higher tip speed ratio, the tip vortices interact much earlier. As a result, $f_r$ surpasses $3f_r$ much closer to the turbine, at $x/D \approx 1$ and remains dominant even beyond 3 rotor diameters. Although $2f_r$ is present, it never becomes dominant, which again indicates that the merging process is fundamentally different for $\lambda = 6$. The strength of wake meandering shows a trend similar to $\lambda = 4.5$ and remains roughly constant throughout the domain of investigation. However, the strength of wake meandering is markedly increased for $\lambda = 6$. An interesting observation can be made if we measure the location where the strength of the wake meandering frequency ($f_{wm}$) surpasses the blade passing frequency $3f_r$. The precise location where it happens depends on $\lambda$. For $\lambda=4.5$ this happens at $x/D \approx 2.2$, while for $\lambda=6$ it occurs at $x/D \approx 1.7$. Interestingly, these distances are close in terms of convective pitch $L_c$ and correspond to $3.15 L_c$ and $3.24 L_c$ respectively. We argue that a distance of roughly $3L_c$ from the rotor plane can be considered as a robust definition of the near wake (where the effect of the individual tip vortices can be felt or beyond which wake meandering can be important) of a turbine independent of the tip speed ratio. The wake meandering frequency, although not dominant, is present in the near wake ($x/L_c<3$), and in fact it exists close to the nacelle in the central region which hints at the fact that the genesis of wake meandering could be related to the shedding of the nacelle or the nacelle/turbine assembly considered together as a porous bluff body. The porosity of the bluff body changes with tip speed ratio, which in turn changes the nature of the vortex shedding from the bluff body. This is explored in more detail in section \ref{wm}.

In the lower plane (experiment $1B$), the frequency spectrum is more broadband and turbulent, as can be seen from fig. \ref{fig:spat_4pt5} and fig. \ref{fig:spat_6}. \NB{As a result, a similar analysis for the lower plane revealed a large number of frequencies of comparable strengths making it difficult to draw firm conclusions hence it is not discussed any further}. The same analysis is also performed for experiment 2 and the strengths of the different frequencies are shown in fig. \ref{fig:evo_top}. The results from experiment $2A$ (\textit{i.e.} at the plane $y=0$) are shown in figs. \ref{fig:evo_top}(a) and \ref{fig:evo_top}(c) for $\lambda=4.5$ and $\lambda=6$ respectively, and they closely resemble fig. \ref{fig:evo_down} despite being different experiments. Note that, for $\lambda=6$ $f_{wm}$ surpasses $3f_r$ exactly at $x/D\approx 1.7$ similarly to fig. \ref{fig:evo_down}(b). The nacelle frequency $f_n$ is important only in the very near field and the tower frequency is rather weak. Fig. \ref{fig:evo_top}(b) and \ref{fig:evo_top}(d) show the strength of frequencies at the offset plane (experiment $2B$). The tower frequency now becomes important and can even be the dominant frequency in the near field. The streamwise evolution of $f_T$ is similar in fig. \ref{fig:evo_top}(b) and (d), showing minimal dependence on $\lambda$. Similarly to fig. \ref{fig:evo_down}, the strength of wake meandering for $\lambda=6$ is higher than $\lambda=4.5$ for both the planes considered, which firmly establishes the fact that the strength of wake meandering is a function of the tip speed ratio. 

\begin{figure}
  \centerline{
  \includegraphics[clip = true, trim = 0 0 20 0 ,width= 0.8\textwidth]{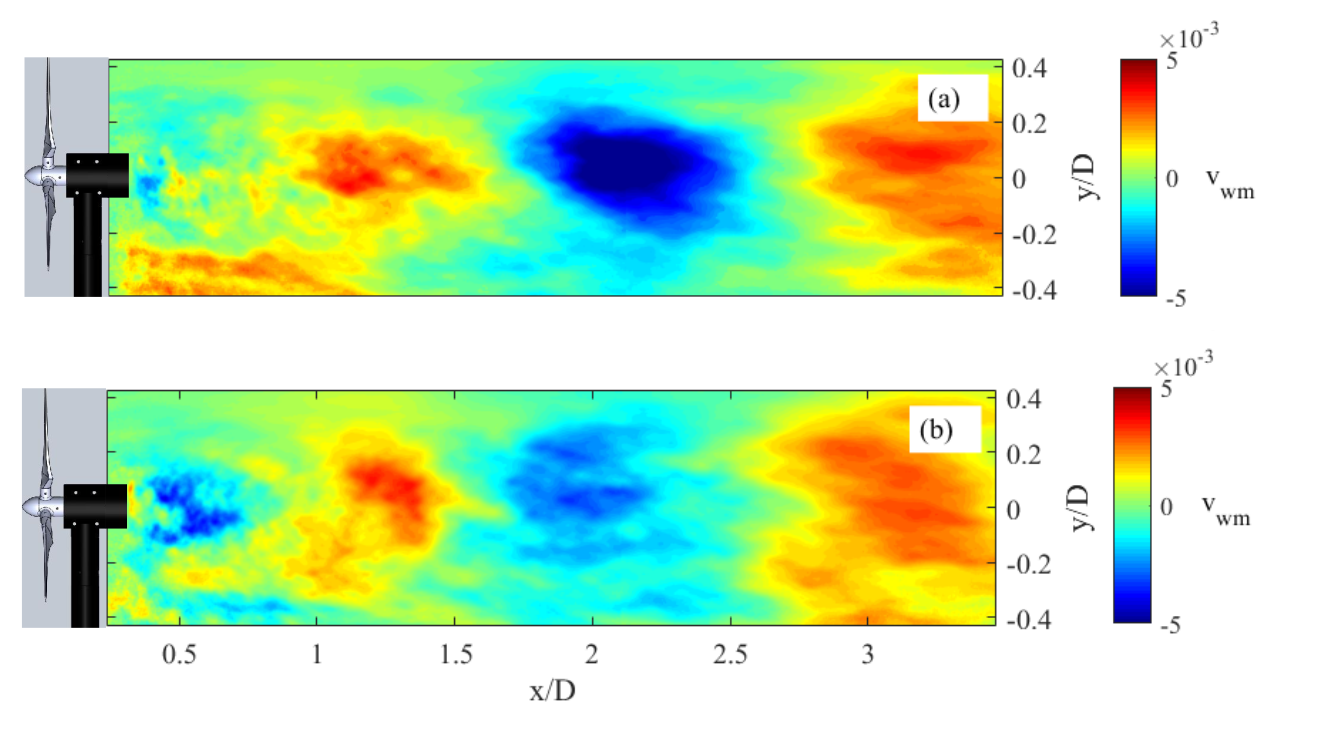}  }
 \caption{Transverse velocity components of the phase-averaged wake meandering modes for (a) $\lambda = 4.5$ and (b) $\lambda = 6$. }
\label{fig:wm}
\end{figure}

\section{Wake meandering}
\label{wm}
Wake meandering has been described as the large scale motion that dominates the `far wake' dynamics of a wind turbine. However, there exists varied opinions about the exact cause of wake meandering \citep{larsen2008wake, espana2011spatial, okulov2014regular, foti2018wake}. 
\NB{Wake meandering has been described as large \emph{intermittent} displacements of the whole wake due to large scale structures in the incoming flow \citep{espana2011spatial}. While \citet{okulov2014regular} reported a well-defined frequency of wake meandering when free stream turbulence was negligible. The wake meandering observed in the present study is of the later type which is induced by the wake-generating body itself, as opposed to inflow turbulence.} Fig. \ref{fig:spat_4pt5} and \ref{fig:spat_6} reveal that the wake meandering frequency is dominant in the central region, within $-0.4D <y<0.4D$. Accordingly, experiment $1C$ focused on this region (see fig. \ref{fig:sch}(a)) to capture the centreline dynamics and the nature of wake meandering. To understand the nature of wake meandering, we utilise phase averaging (see for instance \cite{reynolds1972mechanics,cantwell1983experimental}) based on the frequency of wake meandering observed for the two tip speed ratios. 48 phase bins were used to obtain the phase-averaged flow fields. Thereafter the second Fourier mode of the phase-averaged flow field was obtained and the time varying coefficients were obtained by projecting the phase-averaged modes onto the flow data. More details about this method can be found in \citep{baj2017interscale, biswas2022energy}. \NB{The limited total acquisition time ($\approx$ 10-15 cycles of wake meandering) in experiment $1C$, however, yielded poorly converged modes. Nevertheless, to have a qualitative idea,} the transverse velocity component of the phase-averaged modes are shown in figure \ref{fig:wm}(a-b) for $\lambda=4.5$ and $\lambda=6$ respectively. Note that the modes are qualitatively similar and they resemble vortex shedding from a bluff body of characteristic diameter $D$ which initiates from the vicinity of the nacelle. It indicates that wake meandering is possibly a global shedding mode of the `porous' turbine seeded from the nacelle.

Parallels have been drawn between a rotating turbine and a non-rotating porous disk \citep{lignarolo2016experimental, neunaber2021comparison, vinnes2022far}. It has been shown in several works that the wake characteristics of a porous disk can be similar to that of a turbine, at least in the far field, weakly agreed upon as $x>3D$ \citep{aubrun2013wind, neunaber2021comparison}. The seminal work of \citet{castro1971wake} showed that the Strouhal number of vortex shedding from a porous disk decreases if porosity is reduced and it asymptotes to 0.2 for a solid disk. Note that in a wind turbine, the effective porosity of the rotor disk does change if $\lambda$ is changed. If $\lambda$ is increased, a greater area is swept by the blades in the time taken for a parcel of fluid to convect across the rotor disk, thus increasing the effective blockage, or reducing the porosity. Hence, if wake meandering in a wind turbine wake is related to a global vortex shedding mode, the frequency of wake meandering ought to reduce if porosity is reduced or $\lambda$ is increased. To establish this fact we perform a series of experiments (termed as experiment 3 and 4 in fig. \ref{fig:sch} and table \ref{tab:kd}) which focused on thin strip-like fields of view at different locations in the flow.

\begin{figure}
  \centerline{
  \includegraphics[clip = true, trim = 0 0 0 0 ,width= 1\textwidth]{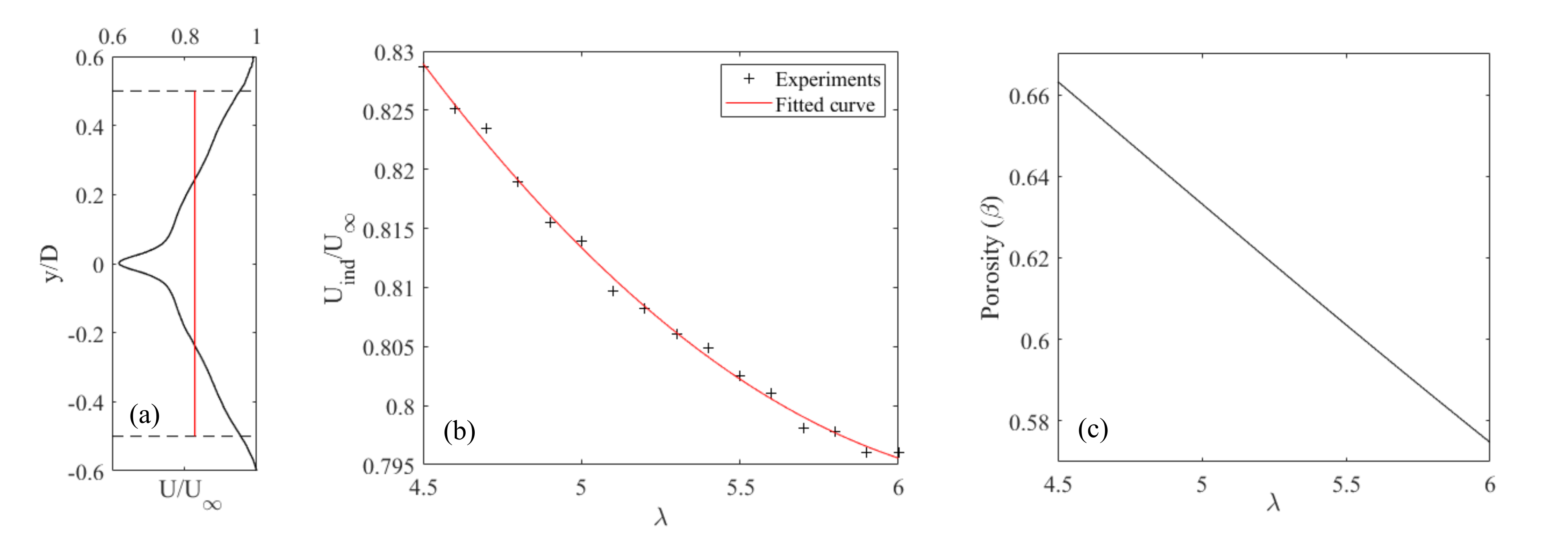}  }
 \caption{(a) Time-averaged velocity profile (solid black line) in front of the rotor disk for $\lambda=4.5$. The red solid line represents the average of the profile within $-0.5<x/D<0.5$ which is denoted as $U_{ind}$. (b) Variation of $U_{ind}$ with $\lambda$. (c) Variation of porosity ($\beta$) with $\lambda$. $\beta$ is defined as the ratio of the unswept open area to the total area of the rotor.}
\label{fig:poro}
\end{figure}

Let us try to estimate the porosity of our model wind turbine as a function of tip speed ratio. First of all, we need a time scale to obtain the area swept by the blades of the turbine at a particular $\lambda$. We can estimate the time scale as the time taken by the flow to pass through the rotor disk. For that we approximate the thickness of the rotor disk as the projected area of the turbine blades (at a plane parallel to $xy$ plane in fig. \ref{fig:sch}) at the root section. Note that the incoming velocity just ahead of the rotor disk is smaller than the free stream velocity due to the induction effect. Hence we obtain the incoming velocity from experiment 4, which considered a thin strip-like field of view just upstream of the rotor (see table \ref{tab:kd} for further details). The black solid curve in fig. \ref{fig:poro}(a) shows the time averaged velocity profile just ahead of the rotor for $\lambda=4.5$. An average of the velocity profile is calculated for $y\in (-0.5D,0.5D)$ which we term as the induction velocity or $U_{ind}$ (shown as red solid line in fig. \ref{fig:poro}(a)). The variation of $U_{ind}$ with $\lambda$ is shown in fig. \ref{fig:poro}(b). With $U_{ind}$ as the velocity scale and the approximate rotor disk's thickness as the length scale we estimate the time taken by the flow to pass through the rotor disk at different $\lambda$. Based on the time scale we obtain the porosity, $\beta$ ($\beta$ is defined as the ratio of the unswept open area to the total area of the rotor disk) of the disk as a function of $\lambda$ and show it in fig. \ref{fig:poro}(c). Note that porosity reduces with $\lambda$ which is consistent with our expectation. 

\begin{figure}
  \centerline{
  \includegraphics[clip = true, trim = 0 0 0 0 ,width= 1.1\textwidth]{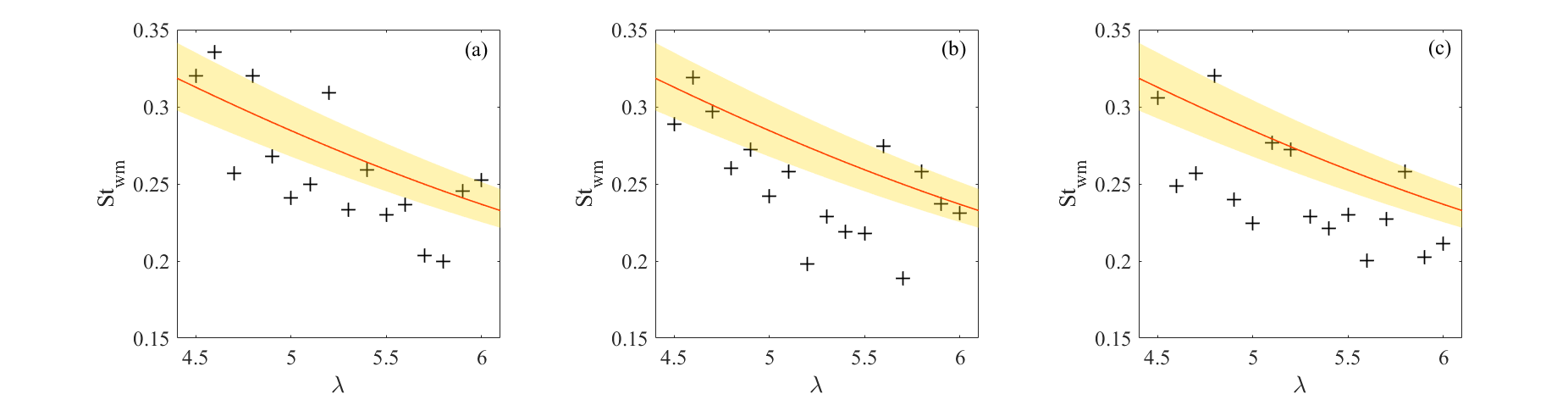}  }
 \caption{Variation of wake meandering strouhal number with $\lambda$ at (a) $x/D=2$, (b) $x/D=3$, and (c) $x/D=5$. `\textbf{\textcolor{red}{---}}' shows the Strouhal number reported by \citet{castro1971wake} for a porous disk at corresponding porosities ($\beta$) with an error margin shown by the yellow shaded region. }
\label{fig:fwm}
\end{figure}

  For a precise measurement of the wake meandering frequency, a large time series of data is required as wake meandering involves rather slow dynamics. A strip FOV was therefore taken which allowed us to obtain a long time series ($\approx 15$ minutes or $\approx 180$ wake meandering cycles) of data considering the storage constraints of the cameras used. The FOVs were centred at three different locations, $x/D = 2$, 3 and 5. Experiments were conducted at 16 tip speed ratios from 4.5 to 6 at steps of 0.1. The Strouhal numbers of wake meandering, $St_{wm}$ (defined based on $D$ and $U_{\infty}$) obtained at different locations are shown in fig. \ref{fig:fwm} with `+' signs. A decreasing trend of wake meandering Strouhal number with $\lambda$ is observed at all three locations. The Strouhal numbers reported by \citet{castro1971wake} for a porous disk at porosities calculated in fig. \ref{fig:poro}(c) are also shown in fig. \ref{fig:fwm} (red line). The yellow shaded region shows the error margin given by \citet{castro1971wake}. The agreement is reasonable considering the approximations that were made to calculate the porosity of the wind turbine at different $\lambda$. This result establishes that wake meandering in a wind turbine wake is related to the global vortex shedding mode of a porous bluff body, the frequency of which depends on $\lambda$ (\textit{i.e.} porosity). Note that a similar decreasing trend of wake meandering frequency with $\lambda$ was reported by \citet{medici2008measurements} at $x/D=1$. \citet{okulov2014regular} reported that wake meandering frequency was a function of operating condition for $1.5<x/D<2.5$, beyond which it was invariant of operating condition. Our results however show that a similar dependence of wake meandering frequency on $\lambda$ persists at least up to $5D$ downstream of the rotor plane.

\section{Conclusion}
We conducted particle image velocimetry (PIV) experiments on the near wake of a lab scale wind turbine model at varying tip speed ratios ($\lambda$). The wind turbine model consisted of a nacelle and a tower to imitate a real scale turbine and to examine the influence of the geometry on the near wake. The freestream turbulence level was low and the wake properties obtained were solely due to the wake generating body. The near field was found to be dominated by the array of coherent tip vortices which appeared to be inhibiting mixing with the outer non turbulent fluid in the immediate near wake, before the tip vortices merged. The merging and the breakdown process of the tip vortices was found to be strongly dependent on $\lambda$. To be precise, for $\lambda=4.5$, a two-step merging process was observed as reported in previous studies \citep{felli2011mechanisms, sherry2013interaction}. For $\lambda=6$ however, there was an earlier and stronger interaction between the tip vortices and the vortices merged directly in an one step process. We defined a length scale termed as the convective pitch ($L_c = \pi D/\lambda$, $D$ being the diameter) that varies with $\lambda$ and is related to the pitch of a helical vortex filament. We proposed that a distance of $3L_c$ from the rotor disk could be used as a robust definition of the immediate near wake (where the tip vortices are not merged) of the turbine irrespective of tip speed ratio. 

Apart from the tip vortices, distinct frequencies associated with the shedding from the tower and nacelle were identified in the near field. The tower frequency is observed over a broad region and it could even be the dominant frequency in the near field for the present experimental condition. Below the nacelle ($y<0$), the interaction of the tip vortices with the tower resulted in an earlier breakdown of the tip vortices and increased levels of turbulence and mixing. Indeed, the tower acted as the major source of asymmetry in the wake also evident by a deflection of the wake centerline towards the tower side ($y<0$). Interestingly, the tower's vortex shedding frequency was observed above the nacelle shear layer as well. Outward bursts of fluid were observed from around the nacelle centreline which are believed to be linked to the oblique nature of vortex shedding from the tower \citep{williamson1996vortex, silvestrini2004direct}. 

\NB{The nacelle frequency was important only very close to the nacelle and was not particularly energetic. However, the nacelle was found to be important in `seeding' wake meandering, indicated by the presence of the wake meandering frequency from relatively close to the nacelle ($x/D<0.5$). A plausible role of the nacelle in aiding wake meandering, combined with the fact that free stream turbulence levels were negligible in the present experiments, uphold wake meandering as a global instability of the turbine wake, the characteristic of which should vary with operating condition. In order to justify this, }separate experiments were performed to calculate $f_{wm}$ at three different streamwise locations, $x/D=2,3$ and $5$ for $4.5\leq \lambda \leq 6$. The Strouhal number of wake meandering was found to decrease when $\lambda$ was increased (or effective porosity was decreased) at all three streamwise stations probed in a similar fashion. Interestingly, a similar decreasing trend of Strouhal number with porosity was observed for vortex shedding behind a porous disk \citep{castro1971wake}. This similarity bolsters the notion that wake meandering is a global instability of the wake generating body, \textit{i.e.} a `porous' turbine with characteristic length scale $D$ and is in contrast to the observation of previous works who found wake meandering frequency to be invariant of operating condition in the far wake \citep{okulov2014regular}. \\

\textbf{Declaration of Interests:} The authors report no conflict of interest.

\bibliography{bib}

\end{document}